\documentclass[12pt]{iopart}
\usepackage{epsfig}
\usepackage{graphicx}

\newcommand{\thickone}{\mbox{$1\!\!1$}}
\newcommand{\thicktwo}{\mbox{$2\!\!2$}}
\begin{document}

\title[Headway oscillations and phase transitions for diffusing particles]{Headway oscillations and phase transitions for diffusing particles with increased velocity}

\author{Marko Woelki\footnote{corresponding author} and Michael Schreckenberg}

\address{Theoretische Physik, Universit\"at Duisburg-Essen, Lotharstr 1, D-47057 Duisburg, Germany}
\ead{woelki@ptt.uni-due.de}
\begin{abstract}
An asymmetric exclusion process with $N$ particles on $L$ sites is
considered where particles can move one or two sites per
infinitesimal time-step. An exact analysis for $N=2$ and a
mean-field theory in comparison with simulations show even/odd
oscillations in the headway distribution of particles. Oscillations
become maximal if particles try to move as far as possible with
regard to their maximum velocity and particle exclusion. A phase
transition separates two density profiles around a generated
perturbation that plays the role of a defect. The matrix-product
ansatz is generalized to obtain the exact solution for finite $N$
and $L$. Thermodynamically, the headway distribution yields the
mean-field result as $N^{-1}\rightarrow 0$ while it is not described
generally by a product measure.
\end{abstract}

\maketitle
\section{Introduction}
Driven-diffusive systems such as the asymmetric simple exclusion
process (ASEP) have been extensively used to model traffic flow
phenomena \cite{schad}. It is defined on a one-dimensional lattice in which
particles can move only in one direction with respect to the
exclusion rule which implies that at most one particle can stay at a
single site. From a theorist's point of view these models are
especially interesting in respect of phase transitions, such as a
jamming and condensation transitions and their solvability at least
for the steady state. Usually, the rather sophisticated models that
claim to reproducing real traffic features are not available for
exact mathematical descriptions. The more important are the
minimalist models that lead to an understanding of the underlying
physics. The ASEP in one dimension with periodic boundary conditions
is fairly simple and has a uniform stationary measure
\cite{Derrida_Altenberg}. There are, however, some generalizations
of the ASEP with periodic boundary conditions that lead to phase
transitions. An example is the ASEP with a single defect particle
that can itself move forward on empty sites and can be overtaken by
normal particles \cite{Mallick}. The defect can be thought of as a
truck that moves in an environment of cars \cite{Derrida_Altenberg}.
Another example is particle disorder: If any of the particles has an
individual fixed hopping rate one might observe a phase transition
from a fluid into a condensed phase \cite{evansrs}. Finally phase
transitions have been studied in asymmetric exclusion models in
which the hopping rate depends on the empty space ahead. These
models can often be related to the zero-range process and the
interactions are normally long-ranged when condensation transitions
appear \cite{evans_1d,hanney}. The ZRP itself allows for an
arbitrary number of particles per site and the single-particle hopping-rate depends only on the occupation on
the departure site \cite{hanney}. This has been generalized to models in
which more than one particle can move. A condition on the hopping
rates has been derived for the steady state to take a simple
factorized form \cite{EMZ}.

In the following section we define a simple traffic model, which is a generalization of
the ASEP in the sense that particles can move one or two sites per infinitesimal time-step.
We find that the system leads to oscillations in the distribution of head-ways which become
maximal when it is impossible to move only one site if there are
more empty sites available. Here the system evolves in special
regions of the configuration space that gives rise to a phase
transition. Although it is a very simple conserving process on a ring with one
species of particles, no overtaking and short-range interactions
it is capable to produce a phase transition and has a non-trivial steady state
that is obtained exactly. We investigated also the process with parallel
update and found the matrix-product stationary state, see
\cite{woelki_jstat}.

\section{The general process}
The general process we are going to investigate is defined on a
one-dimensional lattice with $L$ sites, enumerated $l=1,2,\dots, L$.
Each site $l$ may either be occupied by one particle ($\tau_l=1$) or
it may be empty ($\tau_l=0$). We impose periodic boundary conditions
and let the system evolve in continuous time. Particles can move one
or two sites to the right according to the following rules:
\begin{eqnarray}
\label{def}
100 &\rightarrow & 010, \quad {\rm at \; rate }\; p_1,\nonumber\\
    &\rightarrow & 001, \quad {\rm at \; rate }\; p_2,\nonumber\\
101 &\rightarrow & 011, \quad {\rm at \; rate }\; \beta.
\end{eqnarray}
The parallel-update version of this process has been considered in
\cite{Mukamel}. Note that the total number of particles $N$ is fixed
due to the allowed transitions and boundary conditions. Some simple
cases are already known: For example $p_1=\beta$ was studied in
\cite{klauck} and turned out to have a uniform stationary state. For
$p_2=0$ one finds \cite{klauck,antal} that the weights (for
$\beta>0$) are of the pair-factorized form: $P(\tau_1, \tau_2,
\dots, \tau_L)=\prod_{l=1}^{L}t(\tau_i, \tau_{i+1})$ with some
simple two-site factors $t(\tau_i, \tau_{i+1})$. A mapping onto a
mass- transport model shows that these are the only cases with
factorized steady state (see section \ref{mtm}).
\subsection{Two-particle study}
In the following we consider (\ref{def}) with only two particles on
a ring. The quantity of interest is the un-normalized steady-state weight
$f(m,n)$, denoting that one particle is followed by $m$ and the other by $n$ holes.
Analysis of the master equation shows that these weights obey the following second-order recursion relation
\begin{eqnarray}
\label{rec} \fl f(m,n)=\omega_L(p_1)f(m-1,n)+p_2 f(m-2,n),  \; {\rm
for}\; m\geq n \; {\rm and}\; m+n=L-2\geq 3,
\end{eqnarray}
with the piecewise defined function
\begin{eqnarray}
\omega_{L}(p_{1}) &= \cases{ p_{1}, & for $L$ even,\\
1, & for $L$ odd.}
\end{eqnarray}
One rate can be chosen as 1 and it is convenient to set $p_{1}=1$ to get rid of the even/odd
dependence of the lattice size. Concluding we take for the study of the two-particle case
$\omega_{L}(p_{1}=1)\equiv 1$. In terms of the functions
\begin{equation}
\label{yvonn}
y_{n}:=
\left(\frac{1+\sqrt{1+4p_{2}}}{2}\right)^n +
\left(\frac{1-\sqrt{1+4p_{2}}}{2}\right)^n
\end{equation}
the solution of (\ref{rec}) (for $m \geq n$ and $m\geq 1$) is
\begin{eqnarray}
\label{f(m,n)ende1} f(m,n)&=& \frac{\beta y_{m+n} + (2p_{2}+1-\beta
)y_{m+n-1}}{1+4p_2} + (-1)^{n-1}\frac{(1-\beta
)p_{2}^ny_{m-n}}{1+4p_{2}}.
\end{eqnarray}
One sees that the first term only depends on $n+m$ and therefore it
is constant for given system size. The second term has a pre-factor $(-1)^{n-1}$ which indicates that
in general there are oscillations. Thus the weights $f(m,n)$ depend
on the parity of $n$. The probability for a certain configuration with two particles is given by
$P(m,n)=Z_{m+n+2,2}^{-1}f(m,n)$, where for system size $L$ the normalization $Z_{L,2}=
\sum_{m=0}^{L-2}f(m,L-2-m)$ is:
\begin{eqnarray}
\label{norL2}  \fl Z_{L+1,2}= \left[ \frac{\beta L}{1+4p_{2}} -
\frac{2(1-\beta )(2p_{2}+1)}{(1+4p_{2})^2} \right] y_{L-1} +
\left[\frac{(2p_{2}+1-\beta )L}{1+4p_{2}} - \frac{2p_{2}(1-\beta
)}{(1+4p_{2})^2} \right] y_{L-2}.
\end{eqnarray}
One might think that the feature of oscillations comes from the
presence of a finite number of particles, so we investigate the
thermodynamic limit within a mean-field theory.
\subsection{Mean-Field Theory}
To take into account correlations between consecutive particles we
write down an improved mean-field theory. The quantity of interest
is the probability $P(m)$ to find a headway of $m$ empty sites in
front of a particle. In the context of traffic-flow models this is
referred to as car-oriented mean-field theory (COMF)
\cite{COMF}.\footnote{This formally corresponds to a mean-field
theory in the corresponding mass-transport model (see section
\ref{mtm}) and neglecting correlations between adjacent masses.} The
stationary equations read:
\begin{eqnarray}
\label{MF}
\fl(c+p_{2}s) P(0) &=& \beta P(1) + p_{2} P(2),\nonumber\\
\fl(c+p_2s+\beta)P(1) &=& c P(0) + p_{1}P(2) + p_{2}P(3),\\
\fl(c+p_2s+p_1+p_2)P(m) &=& p_{2}s P(m-2) + c P(m-1) + p_{1}P(m+1) + p_{2}P(m+2),\nonumber\\
&& {\rm for}\; m\geq 2\nonumber,
\end{eqnarray}
with the short-hand notations
\begin{equation}
s:=1-P(0)-P(1) \;{\rm and}\; c:=\beta P(1) + p_{1}s.
\end{equation}
We now introduce the generating function
\begin{equation}
Q(z)=\sum\limits_{m=0}^{\infty} P(m)z^m.
\end{equation}
Summing up $P(m)z^m$ for $m=0\dots \infty$ leads
to a rational expression for $Q(z)$ from which a singularity at $z=1$ can be removed. One obtains
\begin{equation}
\label{Q} Q(z)=\frac{(\beta-p_1-p_2)P(1)z^2-wz-p_2P(0)}{p_2sz^3+(c+p_2s)z^2-(p_1+p_2)z-p_2},
\end{equation}
with $w:=(p_1+p_2)P(0)+p_2P(1)$. A useful check of this equation is
$Q(0)=P(0)$ and $Q(1)=1$. The density $\rho$ in the corresponding
asymmetric exclusion process is
\begin{equation}
\label{rho}
\partial_{z} (zQ(z))|_{z=1} = \sum\limits_{m=0}^{\infty}(m+1)P(m) = \rho^{-1}.
\end{equation}
This gives $P(1)$ in terms of $P(0)$ and $\rho$:
\begin{equation}
\label{P(1)1)}
P(1)=\frac{\left[2p_2(1+\rho)+p_1\right]P(0)-(4p_{2}+p_{1})\rho}
{(\beta-p_{1})(1-\rho)-2p_2}.
\end{equation}
The remaining probabilities can be obtained from $Q(z)$. The flow-density relation
is $J(\rho)=\rho(c+2p_2s)$.

However at this stage already one equation is missing. One needs an
additional relation between $P(1)$ and $P(0)$ to be able to express
everything in terms of the density only. In fact, the missing
relation can be extracted from the generating function. Writing the
numerator of $Q(z)$ in terms of its zeros $z_0^{\pm}$ gives
$(\beta-p_1-p_2)P(1)(z-z_0^{+})(z-z_0^{-})$. The singularity
in the unit circle then has to be removed by $z_0^+$ or $z_0^-$ for $Q(z)$ to be analytic \cite{COMF}.
This leads to the missing relation between $P(1)$ and $P(0)$ \cite{diss}.

\begin{center}
\begin{figure}[!hbt]
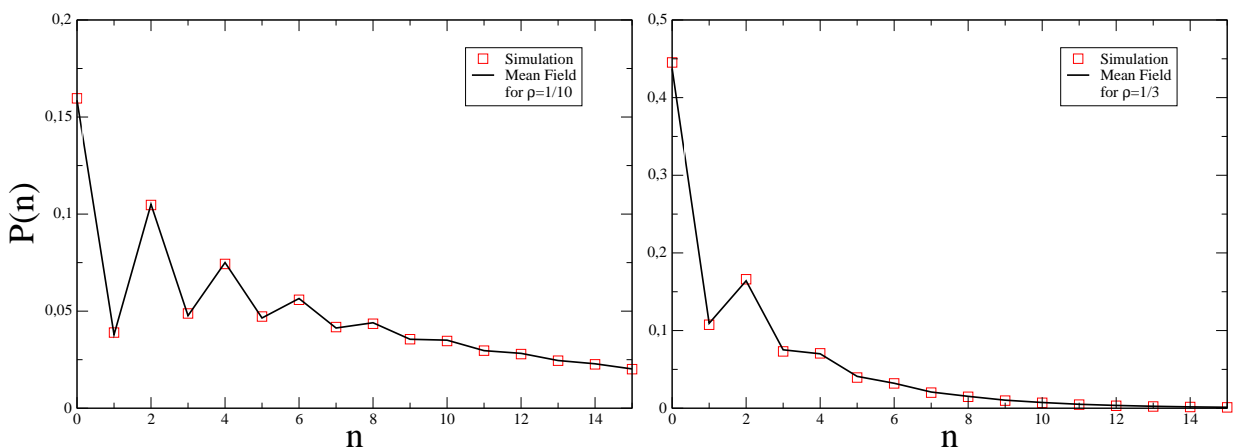

\centerline{\epsfig{figure=p1_01_p2_1_beta_1_01_rho_01,
height=5.8cm, angle=0,
clip=}\epsfig{figure=p1_02_p2_1_beta_1_02_rho_033b, height=5.8cm,
angle=0, clip=}} \caption{Headway distribution $P(n)$ from a
computer simulation with $L=1000$ in comparison with mean field for
$p_2=1$. Left:  $p_1=0.1$, $\beta=1.1$ and $\rho=0.1$. Right:
$p_1=0.2$, $\beta=1.2$ and $\rho=1/3$.} \label{p1p2beta}
\end{figure}
\end{center}

We restrict ourselves here to the case where the coefficient of
$z^2$ in the numerator of $Q(z)$ vanishes, since there the
mean-field predictions can be written in a very compact form and the
main-features that we want to display are contained. Consider
$\beta=p_1+p_2$: a headway's total rate of change is independent of
its length. The condition is weaker than the condition for a
factorized state (every configuration is equally probable if
$p_1=\beta$). Substituting this into (\ref{Q}) and demanding that
the denominator has the same zero as the numerator gives the missing
relation:
\begin{equation}
\label{p1} p_2P(1)^2=p_1P(0)\left(P(0)[1-P(0)]-P(1)\right).
\end{equation}
For $Q(z)$ one gets the simple expression
\begin{equation}
\label{Q2} Q(z)=\frac{pP(0)A^2}{pA^2 - pAz - z^2},
\end{equation}
with $A:=P(0)/P(1)$ and $p:=p_1/p_2$, which can nicely be expanded
to obtain $P(m)$. Figure \ref{p1p2beta} shows the mean-field
distribution for two different choices of parameters in comparison
with computer simulations. While $P(n)$ decays rapidly, one sees the
characteristic even/odd oscillations which are well reproduced by
mean field.
\subsubsection{The Fibonacci case}
\begin{center}
\begin{figure}
\centerline{\epsfig{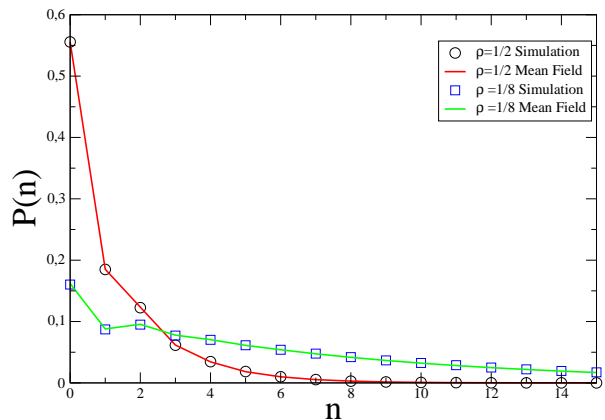}}
\caption{Headway distribution $P(n)$ from mean-field for the Fibonacci
case $p_1=p_2=1$ and $\beta=2$ for $\rho=1/8$ and $1/2$ compared with computer simulations.}
\label{fiB}
\end{figure}
\end{center}

Remarkable is the case $\beta=p_1+p_2=2$ with $p_1=p_2=1$. Then the
probabilities $P(m)$ are given by the Fibonacci numbers:
\begin{equation}
P(m)=P(0)\left(\frac{P(1)}{P(0)}\right)^m F_{m+1}.
\end{equation}
Here one obtains from (\ref{p1}): $P(1)/P(0)=(\sqrt{5-4P(0)}-1)/2$
and relating $P(0)$ to the density gives
\begin{equation}
P(0)=\frac{\rho}{(1+\rho)^2}\frac{5+4\rho-\sqrt{5-4\rho^2}}{2}.\\
\end{equation}
Figure \ref{fiB} shows the headway distribution for the Fibonacci
case.
\subsubsection{The choice $p_1=0$}
In (\ref{Q}) one sees that also $P(1)=0$ reduces the numerators
degree to one.  It has a zero at $z=-p_2(p_1+p_2)^{-1}$. We are
interested in $p_2>0$, so take $p_2=1$ without loss generality.
Demanding that the denominator has the same zero yields (as the only
physical solution) $p_1=0$. The generating function reduces to
$P(0)\left(1+(1-P(0))z^2+(1-P(0))^2z^4+\dots)\right)$. So $P(1)=0$
has the consequence that $P(2n+1)$ vanishes generally and the
process is realized only on the even sublattice in mean field. The
relation to the density is
\begin{equation}
\label{P0} P(0)=\frac{2\rho}{1+\rho},
\end{equation}
and the flow simply reads
\begin{equation}
\label{J} J(\rho)=2\rho\left[1-P(0)\right] =
\frac{2\rho(1-\rho)}{1+\rho}.
\end{equation}
These results are completely equivalent to the usual ASEP where now
particles always move two sites and the density is appropriately
rescaled.

\section{Maximal Oscillations: The choice $p_1=0$}
We consider the process
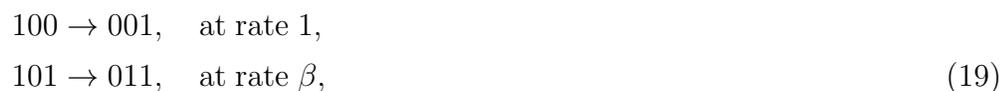
\begin{eqnarray}
\label{def_spec}
100 &\rightarrow & 001, \quad {\rm at \; rate }\; 1,\nonumber\\
101 &\rightarrow & 011, \quad {\rm at \; rate }\; \beta,
\end{eqnarray}
Physically, this is the case where every particle tries to move as
far as possible with regard to his maximum velocity. This is the
limit in which the oscillations in the form of even/odd effects
become maximal. This type of process evolves into special regions of
the configuration space. We try to find the full solution  for
finite $N$ and $L$ from the matrix-product ansatz.

\subsection{Unified solution for finite number of particles and sites} The state of the periodic
system can be expressed by the sequence of headway: $\{n_1, \dots,
n_N\}$. We look for a product state of the form:
\begin{eqnarray}
\label{tens} F(n_{1}, n_{2}, \dots, n_{N}) = {\rm
Tr}\prod\limits_{\mu=1}^{N}\mathcal{G}_{n_{\mu}},
\end{eqnarray}
where $\mathcal{G}_{n_{i}}$ is an operator representing particle
$\mu$ followed by $n_\mu$ holes. It turns out that this ansatz (with
a certain trace-like operation Tr$=$tr$||.||$ to be specified below) yields the correct
steady state, provided that the involved operators fulfill the
following quadratic algebra:
\begin{eqnarray}
\label{algebraMTM} \label{1-}
\mathcal{G}_{2i}\mathcal{G}_{1} & = & \mathcal{G}_{2i+1},\\
\label{2-}
\mathcal{G}_{2i+1}\mathcal{G}_{2j+2} - \mathcal{G}_{2i+1}\mathcal{G}_{2j} & = & 0,\\
\label{3-}
\mathcal{G}_{2i}\mathcal{G}_{2j+2} - \mathcal{G}_{2i}\mathcal{G}_{2j} & = & \beta \mathcal{G}_{2i+2j+2},\\
\label{4-}
\mathcal{G}_{2i}\mathcal{G}_{2j+3} - \mathcal{G}_{2i}\mathcal{G}_{2j+1} & = & \beta \mathcal{G}_{2i+2j+3},\\
\label{5-} \mathcal{G}_{2i+1}\mathcal{G}_{2j+1} & = & 0,
\qquad\quad\quad{\rm for} \;i,j\geq0.
\end{eqnarray}
We just note here that this can be proven by use of the
canceling-mechanism \cite{diss}. The corresponding tagged operators read:
\begin{eqnarray}
\bar{\mathcal{G}}_{0} &=& \mathcal{G}_{0}-\beta \thickone ,\\
\bar{\mathcal{G}}_{2(i+1)} &=& \mathcal{G}_{2(i+1)}+\mathcal{G}_{2i},\\
\bar{\mathcal{G}}_{1} &=&  \mathcal{G}_{1},\\
\bar{\mathcal{G}}_{2i+3} &=& \mathcal{G}_{2i+3}+\mathcal{G}_{2i+1}.
\end{eqnarray}
It is important to emphasize that (\ref{5-}) implies that the system
can not support more than one odd gap. This can be understood
directly from the dynamical rules (\ref{def_spec}). The fact that
the transition $100\rightarrow 010$ is forbidden ($p_1=0$) has the
consequence that the number of odd gaps decreases with time: A
configuration $\mathcal{C}(\dots 1$[any odd number of
0s]$101|\dots$) moves with conditional probability $\beta$ into a
configuration with two odd-valued gaps less (creation of even gaps),
while odd gaps can not emerge. These processes appear until there
remain either no more odd gaps ($L-N$ even) or exactly one odd gap
($L-N$ odd). In the latter case this means physically that the
probability for odd gaps is of order $1/N$ and thus tends
thermodynamically to zero as predicted by mean-field.

One has to be careful with a unified description for an arbitrary number of odd gaps in the system.
The operators $\mathcal{G}$ are mathematical objects that can in our
case be written as matrices whose components are itself matrices (compare \cite{ERS,woelki_jstat}).
Thus it is not obvious how to generalize the trace operation. The
straight-forward generalization to a sum of the traces of the matrices
on the main diagonal can here not be applied: This trace operation
and therewith the weight for certain configurations can incorrectly
give zero. The problem of the trace operation for periodic systems has
previously be pointed out \cite{krebs}.
The matrix relation (\ref{5-}) implies that
$(\mathcal{G}_{2i+1})^{2}=0$ for all $i$. For a unified description
the operators $\mathcal{G}_{2i+1}$ had to be
non-vanishing nilpotent matrices. However for this equation to hold
all the eigenvalues of $\mathcal{G}_{2i+1}$ must be equal to zero.
Therefore also tr $\mathcal{G}_{2i+1}=0$ which implies for example
through relation (\ref{1-}) that also tr
$\mathcal{G}_{2i}\mathcal{U}_{1}=0$ which is untrue. Now this
argument can be used successively to see that for every $N$ the
straight-forward trace operation can not be applied.
First, the algebra (\ref{1-}-\ref{5-}) can be simplified:
\begin{equation}
\label{ans} \mathcal{G}_{2i+1}=\mathcal{E}^{i}\mathcal{A} \;{\rm and
}\; \mathcal{G}_{2i}=\beta\mathcal{E}^{i}\mathcal{D}
\end{equation}
with new operators $\mathcal{E}$ and $\mathcal{D}$. Then  (\ref{1-}-\ref{5-})
reduces to
\begin{eqnarray}
\label{DE}
\mathcal{D}\mathcal{E}&=&\mathcal{D}+\mathcal{E},\\
A\mathcal{E}&=&\mathcal{A},\\
\beta \mathcal{D}\mathcal{A}&=&\mathcal{A},\\
\label{AA} \mathcal{A}^2&=&0.
\end{eqnarray}
Introduce the two by two matrices $\thickone=|1\rangle \langle 1| + |2\rangle \langle 2|$, $\thicktwo=|1\rangle \langle
  2|$ and let further be $\mathcal{E}=E\otimes \thickone$, $\mathcal{D}=D\otimes\thickone$, and $\mathcal{A}=A\otimes\thicktwo$,
with matrices $E$, $D$, and $A$. Then one can interpret the trace
operation in (\ref{tens}) as $F(n_{1}, n_{2}, \dots, n_{N}) = {\rm
tr}||\prod_{\mu=1}^{N}\mathcal{G}_{n_{\mu}}||,$ with the help of the
matrix norm $||M||=$ max$_{i,j}m_{ij}$ to obtain always the correct
weights. A different method is for example a parity-dependent matrix
representation \cite{diss}. However beyond the question of how to
obtain the correct matrix element it is most convenient to consider
both cases separately.

\subsection{Expectation values and phase transition}
For {\it even number of holes} only even gaps occur in the
stationary state. With the particles always making two steps
$(100\rightarrow 001)$ the stationary process is completely
equivalent to the ASEP. All configurations with even-length gaps
have the same stationary weight (the matrices $\mathcal{D}$ and
$\mathcal{E}$ are sufficient to describe the steady state and can be
chosen as numbers). For the normalization we find
\begin{equation}
Z_{L,N}= \frac{L(L-M-1)!}{N!M!}\delta_{L-N,2M}.
\end{equation}
This is easily interpreted combinatorially: For the first particle
one has $L$ possible ways to place it on the lattice. One can then
think of distributing $N-1$ particles and $M$ hole pairs into
$N+M-1$ boxes to obtain the above expression. The flow (from site
$i$) is the expectation value
\begin{equation}
\label{J_corr} J = \langle \tau_i(1-\tau_{i+1})(1-\tau_{i+2})
\rangle + \beta\langle \tau_i(1-\tau_{i+1})\tau_{i+2})\rangle =
\langle\tau_i\rangle - \langle \tau_i\tau_{i+1}\rangle.
\end{equation}
With $\langle \tau_i \rangle =\frac{N}{L}$ and $\langle
\tau_i\tau_{i+1} \rangle =N/L\cdot (N-1)/(L-M-1)$ this yields
asymptotically the mean-field current (\ref{J}). The exact form of
the velocity for finite system size reads
\begin{equation}
v = 2\frac{1-\rho}{1+\rho}\left(1 + \frac{2}{L+N} + \dots +
\frac{2^n}{(L+N)^n} + \dots \right).
\end{equation}
For {\it odd number of holes} exactly one odd gap occurs in the
steady state. If one considers only stationary configurations the
relation (\ref{AA}) becomes redundant. The non-vanishing
steady-state weights from (\ref{tens}) can simply be written as
\begin{equation}
\label{ans_odd} F(2n_1, 2n_2, \dots, 2n_{N}+1) = {\rm tr}\;
\left[\prod\limits_{\mu=1}^{N-1}E^{n_\mu}D\right]E^{n_N}A.
\end{equation}
Here we referred to the particle with the odd gap in front as
particle $N$. Due to the translational invariance this can be done
without loss of generality. The underlying algebra reduces to
\begin{equation}
\label{DEHP} DE=D+E, \quad AE=A, \quad {\rm and }\; \beta DA=A.
\end{equation}
This is the algebra for the ASEP with a single defect particle
\cite{Mallick} for the case $\alpha=1$. This process is defined by
transitions $10\rightarrow 01$ at rate 1, $20\rightarrow 02$ at rate
$\alpha=1$ and $12\rightarrow 21$ at rate $\beta$. In fact, the
stationary states of both processes are completely equivalent. In
our process even gaps of length $2n$ are mapped onto gaps of length
$n$ in the defect ASEP (00 becomes 0). Now consider the single odd
gap with the particle to its right. It can be written in the form 00
00 $\dots $ 01. Again 00 is mapped onto 0 and 01 becomes the defect
2. One can check that under this mapping indeed (\ref{def_spec})
recovers the transitions of the defect ASEP. Note that $\alpha=1$
takes care of the (physically reasonable) fact that in our process particles with an even or
odd gap to their left move forward at the same rate.

\begin{center}
\begin{figure}[!hbt]
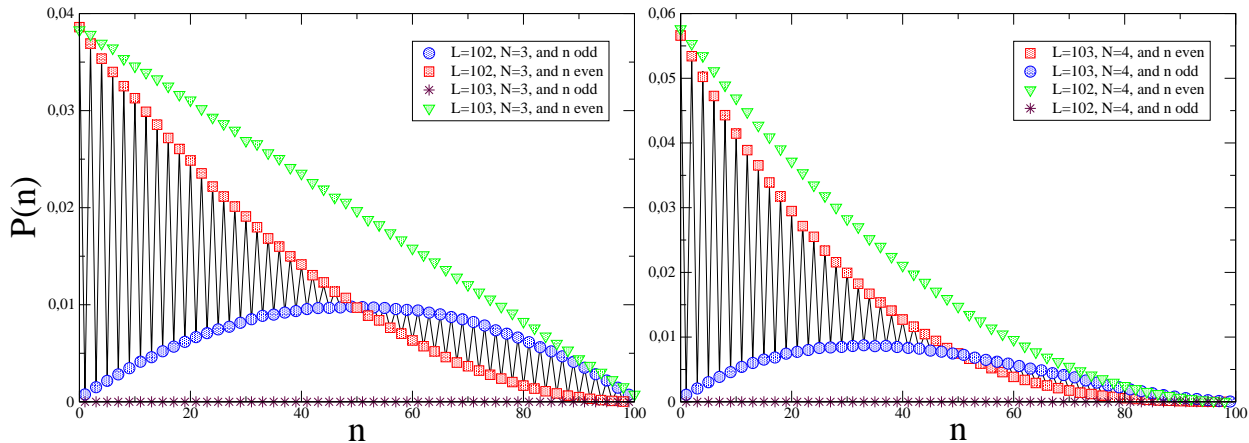

\centerline{\epsfig{figure=L102u103_N3_vergleich_p1_0_beta_1,
height=5.8cm, angle=0,
clip=}\epsfig{figure=L102u103_N4_vergleich_p1_0_beta_1b,
height=5.8cm, angle=0, clip=}} \caption{Headway distribution $P(n)$
for $p_2=1$, $p_1=0$, and $\beta=1$ for $N=3$ and $4$ and $L=102$
and $103$.} \label{fi2}
\end{figure}
\end{center}

As for the defect ASEP the partition function can be calculated:
\begin{equation}
Z_{L,N}=\frac{L}{N}{N+M\choose
N-1}\sum\limits_{m=1}^{\infty}m{N+M-1\choose
N-m}\left(\frac{1-\beta}{\beta}\right)^{m-1}\delta_{L-N, 2M+1}
\end{equation}
From this expression correlation functions can be derived as above.
It turns out that the flow is related to the normalization by
\begin{equation}
J=2\frac{N}{L-2} \frac{Z_{L-2,N}}{Z_{L,N}} +
\beta\frac{L-N}{L-1}\frac{Z_{L-1,N-1}}{Z_{L,N}}.
\end{equation}
We have calculated the finite size expansion for the velocity in the
case $\beta=1$ and obtained
\begin{equation}
v = 2\frac{1-\rho}{1+\rho}\left(1 + \frac{5/2}{L+N} + \dots +
\frac{(1+3^{n+1})/4}{(L+N)^n} + \dots \right).
\end{equation}
The important thing is that the correction is of order $1/(L+N)$
which of course holds also for $\beta \neq 1$.
\\
To summarize in both cases (even and odd number of holes) the
velocity of particles is given by
\begin{equation}
\label{flow}
v=2\frac{1-\rho}{1+\rho}+\mathcal{O}\left(\frac{1}{L+N}\right).
\end{equation}
Just the special form of the correction differs for even and odd
number of holes. Consider now explicitly a finite number of
particles. The weights can be obtained easily from the algebraic
rules. For two particles the weights are of the form $f(2l+1,
2m)=1+l\beta$. For three particles we find $f(2l+1,
2m,2n)=1+(l+n)\beta+\left(\frac{l(l+1)}{2}+ln\right)\beta^2$, and so
on. From these formulae the probabilities $P(n)$ to find a gap of
$n$ sites ($n=0,1, \dots 4$) can easily be calculated. The results
are shown in figure \ref{fi2}. One sees the remarkable change of the
distribution by adding a single empty site to the system. Note that
the probability for an odd gap is of order $1/N$ and thus
independent of $L$. Therefore the qualitative form of holds also for
$L\rightarrow \infty$.
For the thermodynamic limit, one might directly make use of the
density profile \cite{Derrida_Altenberg,Mallick}.

\begin{center}
\begin{figure}[!hbt]
\centerline{\epsfig{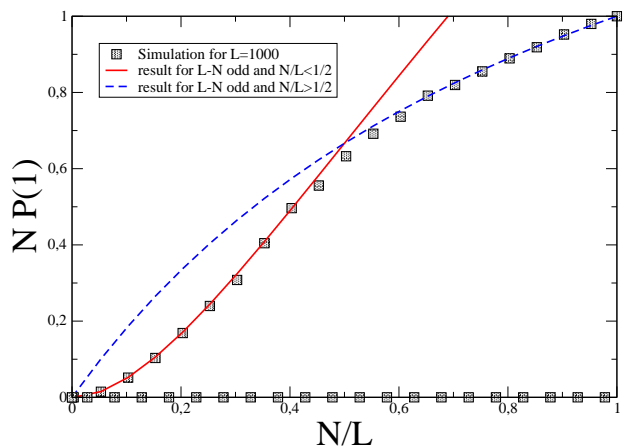}} \caption{$P(1)$ versus $\rho$ for $\beta=2/3$. See text for details.}
\label{pt}
\end{figure}
\end{center}

It is interesting that the single `excess hole' created by the
dynamics leads to effects also in the thermodynamic limit. Although
the probability $P(n)$ as predicted by mean field is
thermodynamically exact, the system reaches no product-measure
steady state, since around the defect a nontrivial density profile
is formed. Beyond that a phase transition takes place equivalent to
a transition in the defect ASEP \cite{Mallick}. In terms of the
different density ($\rho$ in the (defect) ASEP becomes here
$2\rho/(1+\rho)$) the critical density is $\rho_c=\beta/(2-\beta)$.
In the following let us in analogy refer to the $01$-pair as the
defect. Since we have $\alpha=1$ the phase diagram of the defect
ASEP reduces to a single line.
\begin{itemize}
\item For $\rho>\rho_c$ the defect behaves as the other particles. In
front of the defect the density profile decreases exponentially to
its
bulk value. The density behind is constant.\\
\item For $\rho<\rho_c$ the defect is similar to a second-class
particle \cite{Derrida_2class} that lowers the average speed of the
other particles. The density profile decays algebraically to the
bulk value. Behind the defect the density is decreased and the
profile increases in the same way to its bulk value as in front. The
profile is the limit of a shock profile with equal densities to the
left and right.
\end{itemize}
From the relation to the defect ASEP one can obtain the
probabilities $P(2n+1)$ for odd headway. For example $P(1)$ is
related to the probability $\rho_-$ in \cite{Mallick} to find a
particle directly behind the defect. Since in our process the defect
can be any of the $N$ particles one has
\begin{equation}
\label{P1}
[P(1)](\rho)=\cases{\frac{4\rho^2}{\beta N(1+\rho)^2},&for $\rho<\rho_c$,\\\frac{2\rho}{N(1+\rho)},&for $\rho>\rho_c$.}
\end{equation}
Figure \ref{pt} shows $P(1)$ scaled with $N$ versus the density for
$\beta=2/3$, so that the phase transition happens at $\rho_c=1/2$.
Depicted are the analytic formulae from (\ref{P1}) together with a
computer simulation for $L=1000$ with $N$ increased in steps of
$\Delta N=25$.

\subsection{Relation with generalized zero-range processes}
\label{mtm} In the previous section we expressed the steady state of
the system through the set of gaps between particles. This formally
corresponds to a mapping onto a model in which a site occupied by
particle $\mu$ becomes site $\mu$ and the gap $n_{\mu-1}$ to the
left becomes the `mass' on site $\mu$. The ansatz (\ref{tens}) then
becomes site-oriented. The corresponding process comprises $N$ sites
and $M:=L-N$ particles. In the process obtained from
(\ref{def_spec}) by this mapping one or two particles may leave a
certain site with rates:
\begin{eqnarray}
\label{mass_spec} \gamma(l|m)=\cases{
1, & for $l=2, m>1$,\\
\beta, & for $l=1, m=1$.}
\end{eqnarray}
However this is a special case of the class of generalized
zero-range processes introduced by Evans, Majumdar and Zia
\cite{EMZ} and referred to as `mass-transport models' (MTM). They
derived a necessary and sufficient condition for the steady state to
factorize, i.e.\ $P(m_1, m_2, \dots, m_N) \propto
\prod_{\nu=1}^{N}f(m_{\nu})$. The condition on the chipping
functions $\gamma$ reads \cite{EMZ,ZEM}:
\begin{equation}
\label{fact} \gamma(l|m)=\frac{w(l)f(m-l)}{f(m)}, {\rm for }\; l=1,
\dots, m
\end{equation}
where $w(l)$ is an arbitrary non-negative function of $l$. For the
process (\ref{mass_spec}) the situation is slightly more special.
Here we have a factorized steady state only for even mass and thus
it is not predicted by (\ref{fact}). In our rather singular case,
involving vanishing single-site weights, (\ref{fact}) is not
fulfilled, as is easily seen for $\gamma(1|m)$. To understand this,
let us consider a slightly more general case. Assume general
single-site weights that vanish for odd masses as it is the case for
(\ref{mass_spec}). Considering the master equation shows that the
restrictions for the chipping function to recognize the factorized
state for $M$ even, are:
\begin{eqnarray}
\label{genMTM} \gamma(l|m+l) = \cases{{\rm free } ,& for $m+l$ odd,\\ 0, & for  $l,m$ both odd,\\
\frac{x(l)f(m)}{f(m+l)}, & for $l,m$ both even.}
\end{eqnarray}
However the choice for $f(m)$ in turn implies that the total mass $M=\sum_{i=1}^{L}m_{i}$ is even.
Otherwise the normalization
\begin{equation}\label{norm}
Z_{N,M} = \sum\limits_{\{m_{i}\}}\delta \left(
M-\sum_{j}m_{j}\right)\prod\limits_{i=1}^{N}f(m_{i})
\end{equation}
would vanish. The solution can be used to obtain new solvable
models: The process (\ref{mass_spec}) leads to a matrix-product
state for the choice of $M$ for which the system can not reach the
factorized state. This suggests that the general model
(\ref{genMTM}) with odd particle number (defined trough the weights
for even $M$) may also lead to a matrix-product state by special
choice of the free parameter in (\ref{genMTM}). Of course these
arguments can be generalized to other choices of vanishing weights
\cite{diss}. Beyond that, the single-site mass-distribution should
equal thermodynamically the result obtained from the case where it
is factorizable, since in the infinite system a local perturbation
changes the density profile but not the single-site distributions.
This way one can obtain the exact distributions also for cases where the steady state has not generally a product measure.\\
Some focus has recently be attended to two-species zero-range
processes and conditions for a factorized steady state
\cite{hanney}. We consider model parameters that violate this
condition. Denote the number of first-class particles on site $l$ by
$n_l$ and the number of second-class particles as $m_l$. The rates
$u(n,m)$ and $v(m,n)$ at which first and second-class particles move
are taken as
\begin{eqnarray}
u(n,m) & = & 1, \quad\quad\quad\quad\quad {}{\rm  for }\; n\geq 1,\\
v(m,n) &=  & \beta \left[1-\theta(n)\right], \quad{\rm for }\; m
\geq 1.
\end{eqnarray}
The motivation for this choice is simply that the steady state of the MTM
(and equivalently (\ref{def_spec})) corresponds to the case of a single 2-particle here
(pairs of particles in the mass-transport model are mapped onto 1-particles in the ZRP, and
the single excess particle is mapped onto the 2-particle). The resulting
algebra then also becomes a consequence of (\ref{DEHP}).

\section{Conclusion and outlook}
To summarize, a simple traffic model that generalizes the ASEP with
periodic boundaries has been considered. In this continuous-time
process, particles can move one or two sites to the right. This
mimics a larger maximum velocity of cars that is typical for
discrete traffic models \cite{nasch}. Some preliminary results have
already been published \cite{woelki_tgf}. We considered choices of
hop rates that are not appropriate for modeling traffic but are of
theoretical interest. A mean-field theory in comparison with
simulations showed that the headway distribution of particles show
even/odd oscillations. We considered a special choice of the
parameters (see (\ref{def_spec})) for which the oscillations become
maximal. The matrix-product ansatz was generalized to obtain the
exact solution for finite number $N$ of particles and sites $L$.
Thermodynamically a crucial phase transition appears: Particles move
at their desired maximum velocity $v_{max}$ which leads to the
formation of headway being a multiple of $v_{max}$. If there remains
smaller headway, then it moves through the system against the
direction of motion as a sort of defect. This defect (being
somewhere in the system) changes locally the density profile but not
the headway distribution of a single particle. Therefore the system
is not described by a product measure although the asymptotic
headway distribution is given by the mean-field result. Instead the
solution is the matrix-product state for the ASEP with a single
defect \cite{Mallick}. Note that the process with parallel dynamics
leads to a similar stationary state with additional attraction
between particles and hole-pairs and can also be solved \cite{woelki_jstat}.\\
We established a relation between non-ergodic exclusion processes
with higher velocities, generalized zero-range processes (ZRP) and
defect systems. This way one can calculate exact quantities without
knowing the exact density profile. For future work it is interesting
to investigate the connection between systems with creation and
annihilation of `defects' and generalized ZRP to be able to handle
ergodic dynamics without parity dependence.\\
The condition for a simple factorized state in the ASEP considered
here is that the rate at which a particle moves a certain number of
sites is independent on the headway. This condition holds also for
parallel dynamics \cite{wss} and higher $v_{max}$. We investigated a
slightly more general parameter line, where the rate at which a
particle can change its headway does not depend on its length which
allows for oscillations in the headway distribution. The mean-field
assumption leads to a remarkable agreement with computer
simulations. We pointed out the Fibonacci case whose corresponding
mass-transport model (on two sites) has a factorized steady state.
The single-site weights become Fibonacci numbers $s(n)=F_{n+1}$. The
recursion relation (\ref{rec}) becomes $f(m,n)=f(m-1,n)+f(m-2,n)$,
for $m\geq 2$, which can be expressed as a matrix-product state
$f(m,n)= {\rm tr}\;\left(DE^mDE^n\right)$ with $DEE=DE+D$. However
neither the factorization nor the matrix recursion hold for more
than two particles. So this case remains unsolved. A comparable good
agreement with mean field has been observed previously for the ASEP
with shuffled update \cite{woelki_shuffled} where also the
two-particle state is factorizable. It would be interesting to find
out whether the mean-field result in these models is generally exact
although the process has not a product measure. Beyond that, the
connection between the solvability for two particles and $N$
particles is still an open problem. Here one might be able to profit
from knowledge in
equilibrium statistical mechanics \cite{baxter}.\\
We would like to thank A. Schadschneider for helpful discussions.

\end{document}